\documentclass[prl,twocolumn,showpacs,preprintnumbers,superscriptaddress,amsmath,amssymb]{revtex4}

\usepackage{epsf,color,graphicx}

\usepackage[dvips]{epsfig}

\begin{document}


\title{Polariton laser using single micropillar GaAs-GaAlAs semiconductor cavities}



\author{Daniele Bajoni}
\author{Pascale Senellart}
\author{Esther Wertz}
\author{Isabelle Sagnes}
\author{Audrey Miard}
\author{Aristide Lema\^{i}tre}
\author{Jacqueline Bloch}\email[]{jacqueline.bloch@lpn.cnrs.fr}
\affiliation{CNRS-Laboratoire de Photonique et Nanostructures, Route
de Nozay, 91460 Marcoussis, France}

\date{\today}

\begin{abstract}
Polariton lasing is demonstrated on the zero dimensional states of
single GaAs/GaAlAs micropillar cavities. Under non resonant
excitation, the measured polariton ground state occupancy is found
to be as large as $10^{4}$. Changing the spatial excitation
conditions, competition between several polariton lasing modes is
observed, ruling out Bose-Einstein condensation. When the polariton
state occupancy increases, the emission blueshift is the signature
of self-interaction within the half-light half-matter polariton
lasing mode.
\end{abstract}

\pacs{78.55.Et, 71.36.+c, 78.45.+h}

\maketitle

Boson statistics can lead to massive occupation of a single quantum
state and trigger final state stimulation. This stimulation is
responsible for the bright coherent emission of light in a laser.
Another fascinating property of massive bosons in thermal
equilibrium is their ability to accumulate in the lowest energy
state under a given critical temperature. First predicted in
1925,\cite{Einstein} the experimental observation of Bose Einstein
condensation was achieved in the mid 1990s for ultra-cold
atoms.\cite{Anderson1995,Ketterle1995} Demonstrating such bosonic
effects with matter waves in a solid state system is very
interesting both from fundamental point of view but also for
applications since it could provide a new source of coherent light.
Cavity polaritons are an example of quasi-particles behaving as
bosons at low density.\cite{Livrekavokin,Keeling2007} They are the
exciton-photon mixed quasi-particles arising from the strong
coupling regime between quantum well (QW) excitons and a resonant
optical cavity mode. Because of their very small effective mass
($10^{-8}$ times that of the hydrogen atom) cavity polaritons are
expected to condensate at unusually high temperatures (up to room
temperature in wide band gap microcavities).\cite{Malpuech2002}
These last years, massive occupation of a polariton state has been
observed in semiconductor two-dimensional (2D) cavities and
attributed to Bose Einstein condensation\cite{Deng2002,Kasprzak2006}
or to polariton lasing.\cite{Christopoulos2007} More recently,
polariton condensation has been claimed in a localized energy trap
\cite{Snoke2007} where the trap dimensions are sufficiently large
for the system to present a 2D continuum of polariton states. In
these experiments, the clear distinction of a thermodynamic phase
transition (Bose Einstein condensation) from a kinetic stimulated
scattering (polariton lasing) is still debated.

In this letter, we demonstrate polariton lasing in micrometric sized
GaAs/GaAlAs micropillar cavities. In such zero-dimensional (0D)
cavities, polariton states are confined in all directions and
present a well defined discretized energy
spectrum.\cite{Gerard,Panzarini} The absence of translation
invariance lifts the wave-vector conservation selection-rules in
polariton scatterings. In GaAs 2D microcavities, these selection
rules are responsible  for inefficient polariton-phonon or
polariton-polariton scattering, preventing the build-up of a large
occupancy in the lower energy
states.\cite{Senellart2000,Tartakovskii2000,Butte2002,bajoni} In
this work, we show that polariton scattering is very efficient in
micropillar cavities. Under non resonant excitation, a threshold
corresponding to a measured occupation factor equal to unity is
observed, followed by a massive occupation of the lowest energy
polariton state. At higher excitation power, the progressive
transition from the strong to the weak coupling regime is evidenced
with the onset of conventional photon lasing.

Moving the excitation spot toward the micropillar edge, non-linear
emission can be triggered on higher energy polariton states. Such
behavior is characteristic of a polariton laser with competing
stimulated scattering toward several polariton modes. It rules out
Bose-Einstein condensation where only massive occupation of the
ground state is expected. Finally the spectral blueshift of the
polariton laser line is shown to be induced by the polariton
self-interaction within the lasing mode. This experiment is the
first demonstration of a solid state matter-wave laser on 0D states.

\begin{figure}[]
\includegraphics[width= 0.8\columnwidth]{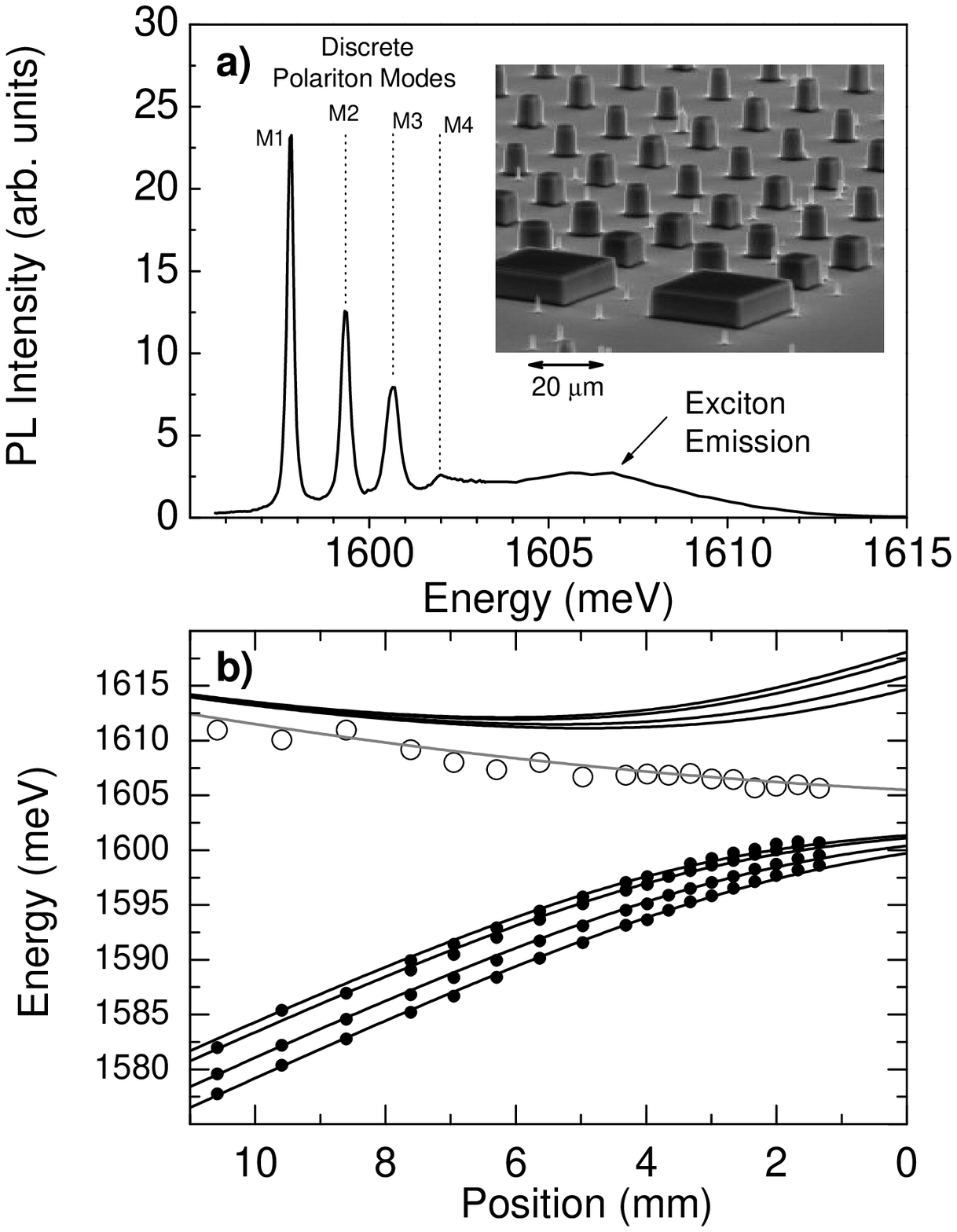}
\caption{(a) PL spectrum measured on a single 4 $\mu$m diameter
micropillar; inset: scanning electron micrography of a micropillar
array; (b) (black circles) Emission energy of the discrete polariton
PL lines and (open circles) of the exciton line measured on several
4 $\mu$m diameter micropillars along the cavity wedge, (thick black
lines) calculated energy of the 0D polariton modes. T = 10
K.}\label{Fig1}
\end{figure}

Our sample, grown by molecular beam epitaxy, consists in a
$\lambda/2$ Ga$_{0.05}$Al$_{0.95}$As cavity surrounded by two
Ga$_{0.05}$Al$_{0.95}$As/Ga$_{0.80}$Al$_{0.20}$As Bragg mirrors with
26 and 30 pairs in the top and bottom mirrors respectively. Three
sets of four 7 nm GaAs QWs are inserted at the antinodes of the
cavity mode electromagnetic field: one set is located at the center
of the cavity layer and the two others at the first antinode in each
mirror.\cite{Bloch1998} A wedge in the layer thickness allows
continuous tuning of the cavity mode energy $E_{C}$ with respect to
the QW exciton energy ($E_{X}$). The exciton-photon detuning is
defined as $\delta = E_{C}-E_{X}$. 20 to 2 $\mu$m size square and
circular micropillars were fabricated along the wafer using electron
beam lithography and reactive ion etching (see inset of Fig.
\ref{Fig1}(a)). Photoluminescence (PL) experiments are performed on
single micropillars using a cw Ti:Saph laser focused onto a 3 $\mu$m
diameter spot with a microscope objective. For excitation powers
exceeding 1 mW, the laser beam is chopped using an acousto-optic
modulator with 1\% duty cycle at 10 kHz. The emission is collected
through the same objective, spectrally dispersed and detected with a
nitrogen cooled CCD camera. The sample is maintained at 10 K in a
cold finger cryostat. The laser is tuned to the first reflectivity
minimum on the high energy side of the mirror stop-band (around 740
nm), typically 80 meV above the exciton resonance.

In micrometer sized pillar cavities, photons are confined along all
directions: vertically by the Bragg mirrors and laterally by the
index of refraction contrast between air and semiconductor. As a
result, micropillars exhibit discrete 0D photon modes.\cite{Gerard}
In the strong coupling regime, polaritons come from the mixing
between each of these 0D photon modes and the QW
excitons.\cite{Panzarini} Fig. \ref{Fig1}(a) presents a PL spectrum
measured on a single 4 $\mu m$ diameter circular micropillar. The
emission energies measured on micropillars of identical diameter
along the cavity wedge are summarized in fig.\ref{Fig1}(b). The
emission spectrum in fig. \ref{Fig1}(a) presents several discrete
emission lines on the low energy side of the exciton line centered
around 1607 meV. The energy of these discrete lines strongly varies
with the layer thickness: they are attributed to 0D photon modes.
For large negative detuning, the 130 $\mu$eV linewidth of these
optical modes corresponds to a quality factor of 12000. Each of
these photon modes presents the anticrossing\cite{Bloch1997,
Gutbrod1998} with the exciton line, characteristic of the strong
coupling regime. The 0D polariton energies can be fitted using a 15
meV Rabi splitting (as in the planar cavity). Note that contrary to
2D cavities, the uncoupled exciton line is observed because in-plane
exciton emission is extracted through the pillar side. Since the
exciton line is broadened by strain relaxation in the etching
process, the upper polariton state could not be resolved. Further
evidence of the strong coupling regime is obtained by observing the
exciton-photon anticrossing on a single micropillar using
temperature tuning.\cite{Fisher1995}

\begin{figure}[]
\includegraphics[width= 0.8\columnwidth]{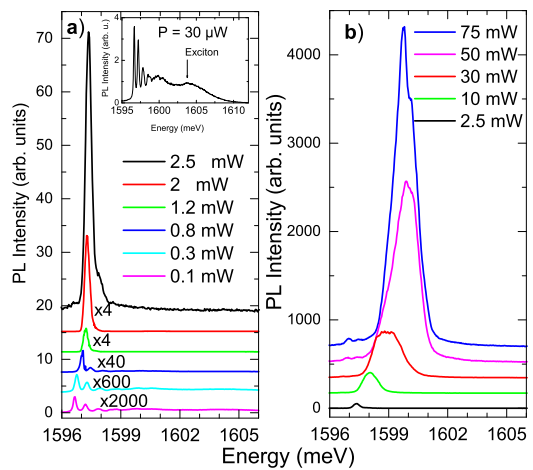}
\caption{(Color online) Emission spectra measured for several
excitation powers in a single 6 $\mu$m diameter micropillar.(a)
corresponds to the polariton lasing regime. The inset shows the
emission spectrum measured at a low excitation power of 30 $\mu$W.
(b) corresponds to the loss of the strong coupling regime and the
onset of photon lasing. T = 10 K.}\label{Fig2}
\end{figure}

PL spectra taken for increasing pump powers $P$ on a 6 $\mu$m
circular pillar are shown in fig. \ref{Fig2}. A spectrum taken at
very low power is shown in the inset: several discrete polariton
modes emit on the low energy side of the exciton line. The polariton
ground state corresponds to a $50\%$ exciton $50\%$ photon mixed
state (measured detuning $\delta = 0$ meV). Two excitation regimes
can be distinguished. For $P> 5 mW$ (fig.\ref{Fig2}(b)), the
emission undergoes a pronounced blueshift and broadening. The
density of electron-hole pairs per QW injected for  $P=5 mW$ is
estimated to be around $10^{10}$ cm$^{-2}$ per QW, reaching the
exciton screening density. In this excitation range, the strong
coupling regime is progressively
screened\cite{Houdre95,Kira97,Pau97} and eventually the system
enters the weak coupling regime, with emission of uncorrelated
electron-hole pairs through the cavity modes. The strong coupling
regime saturation is observed for excitation densities consistent
with previous reports in 2D samples\cite{Houdre95} or in
aluminium-oxide-aperture nanocavities\cite{Lee}. Above 40 mW, a
threshold is observed due to the onset of conventional photon lasing
on the 0D photon modes\cite{footnote}.

\begin{figure}[]
\includegraphics[width=7 cm]{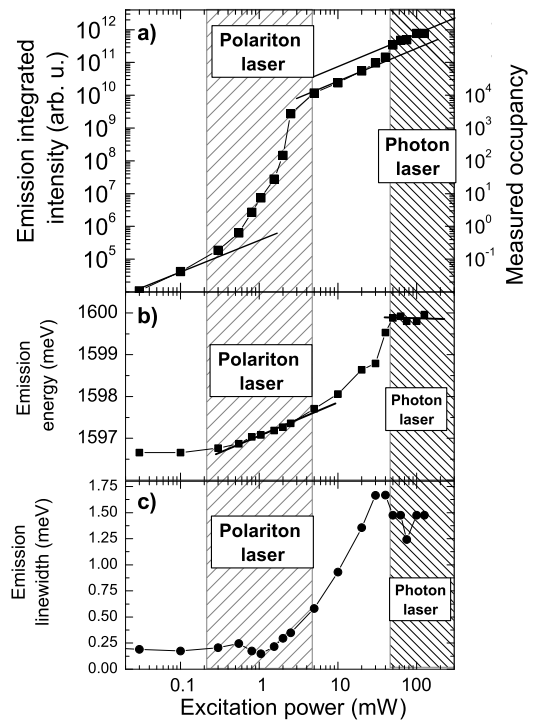}
\caption{(a) Integrated intensity and measured occupancy , (b)
emission energy and (c) emission linewidth measured on the lowest
energy emission line as a function of the excitation power; dashed
areas highlight the excitation range for polariton lasing or photon
lasing.}\label{Fig3}
\end{figure}

Let us now concentrate on the strong coupling regime ($P<5$ mW,
fig.\ref{Fig2}(a)). The emission behavior is marked by a sharp
nonlinear increase of the ground state PL intensity. As summarized
in Fig. \ref{Fig3}(a), when P varies from 0.3 mW up to 3 mW, the
integrated intensity increases by four orders of magnitude. The
occupancy N of the lowest energy polariton state can be
experimentally estimated using: $N = I_{PL}*
\tau_{cav}/[\alpha^{2}*E]$, where $\tau_{cav}= 3\, ps$ is the cavity
photon lifetime, $I_{PL}$  is the power emitted by the considered
polariton mode, E its emission energy and $\alpha$ its exciton
part\cite{Senellart2000}. Directly relating the emission intensity
to an actual polariton population may not always be possible since
emission at the polariton energy could also come from correlated
electron-hole pairs, as reported for bare excitons\cite{Chatterjee}.
Nevertheless this analysis is valid at low density and low
temperature, the regime where polariton non-linearities are observed
in the present work. As shown in Fig. \ref{Fig3}(a), the onset of
the polaritonic non-linearity occurs when the measured polariton
occupancy exceeds unity. Above threshold, the scattering of excitons
toward the lowest energy polariton state is stimulated, leading to
the formation of a macroscopically occupied polariton state. The
polariton population of the lower energy state increases up to
10$^4$.

The polariton linewidth amounts to $0.2$ meV at low excitation
power, and slightly broadens with the onset of polariton-polariton
interaction. At threshold, associated to the build-up of the large
polariton occupancy, a spectral narrowing, down to $\sim$ 0.15 meV,
shows that the coherence time becomes longer than the radiative
lifetime of single polaritons.

Polariton stimulated scattering is obtained on micropillars with
diameter down to 2 $\mu$m and for detunings down to $\delta =-$20
meV. Working at higher temperatures, polariton lasing occurs up to
45 K. Above 50 K or for very large negative detunings, only the
second non-linearity associated to photon lasing is observed. Let us
underline that in the unpatterned 2D sample, only conventional
lasing was achieved\cite{bajoni}. Reducing the cavity dimensionality
is the key step to achieve polariton quantum degeneracy under
non-resonant excitation.

\begin{figure}[]
\includegraphics[width=7 cm]{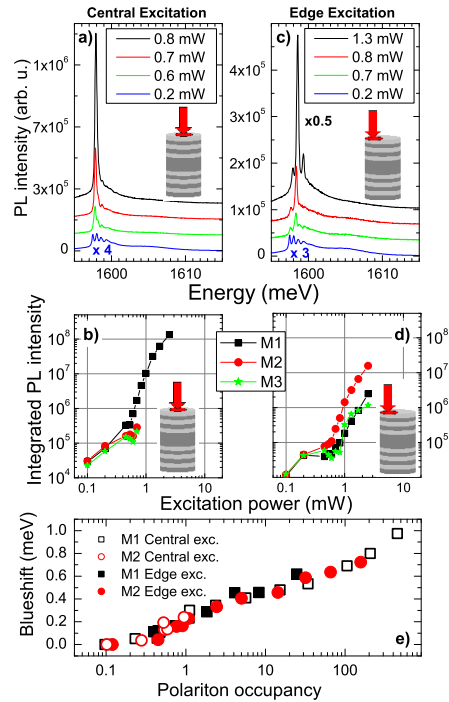}
\caption{(Color online) (a) Emission spectra measured on a single 6
$\mu$m diameter micropillar for several excitation powers with a
laser spot well centered on the pillar surface, (b) measured
integrated intensity of the three lowest energy polariton mode as a
function of the excitation power in this centered excitation
geometry. (c) and (d): same as (a) and (b) but with an edge
excitation geometry as schematically indicated in the figure. (e)
Spectral blueshift of the polariton lasing modes as a function of
their occupancy. T = 10 K.}\label{Fig4}
\end{figure}

Fig. \ref{Fig4} presents PL measurements on a 6 $\mu$m micropillar
(with $\delta =-$0.3 meV for the lowest polariton mode) using two
different excitation geometries. On the left part of Fig.\ref{Fig4},
the laser spot is centered on the micropillar surface. As described
above, stimulated scattering occurs toward the lowest energy mode
(named M1). The data of the right part are recorded with the
excitation spot shifted toward the micropillar edge, as
schematically indicated in the figure. Under this excitation
geometry, stimulated scattering is observed toward the first excited
polariton states (named M2). Further increasing the excitation
power, competition between polariton modes also triggers stimulated
scattering toward M1 and M3. This experiment demonstrates that the
polaritonic non-linearity can not be interpreted in terms of a
thermodynamic phase transition (analogous to Bose-Einstein
condensation) because in this framework, the largest occupancy is
always expected on the system ground-state. In the present
experiment, the non-linearities are not governed  by thermodynamics
but by the kinetics of the scattering process toward the low energy
polariton states. Such bosonic stimulation of polariton scattering
has been named "polariton laser"\cite{Imamoglu96,Shelykh2003} in
analogy to the atom laser\cite{Wiseman}. Using this now well
accepted name, one must keep in mind that the stimulation mechanism
is very different from that in a conventional photon laser: it is
not the emission of radiation that is amplified but a scattering
mechanism, following the non-resonant excitation. As in a
conventional photon laser, multi-mode polariton lasing can be
triggered depending on the excitation condition. The electromagnetic
field of the lowest energy mode (named HE$_{11}$, see
ref.\onlinecite{Panzarini}) presents an antinode at the micropillar
center and decays at the edge. It is therefore favored when the
center of the pillar is excited. On the contrary, the second
polariton line gathering three degenerate modes within the mode
linewidth (EH$_{01}$, HE$_{21}$ and HE$_{01}$), is favored under
edge excitation since these modes present an antinode at the
periphery. Thus the pump excitation geometry triggers stimulated
scattering toward polariton modes with matching field spatial
distribution.

Contrary to conventional photon lasers, the lasing mode of a
polariton laser is macroscopically occupied with half-matter
half-light bosons i.e. with interacting bosons. Exciton-exciton
interaction and the resulting exciton blueshift has been extensively
studied in the 80's\cite{Peyghambarian,Schmitt-Rink86} and recently
revisited within the framework of excitonic
polaritons\cite{Ciuti2000}. Polariton-polariton interaction
originates from coulomb interaction between the fermionic
constituents (electron and hole) of their exciton part. As the
occupancy of the polariton states increases, self-interaction within
the lasing mode induces a spectral blueshift of the
emission\cite{Ciuti}. Fig. \ref{Fig4}(e) summarizes the spectral
blueshift of M1 and M2 both under central and edge excitation
conditions. The blueshifts are plotted as a function of the
polariton occupancy, obtained by normalizing the emission intensity
by the intensity at threshold. M3 is not reported because the
contribution from HE$_{12}$, at higher energy than EH$_{11}$ and
HE$_{41}$\cite{Panzarini} could not be correctly deconvoluted. The
curves in  fig.\ref{Fig4}(e) are strikingly identical regardless of
the mode number or the excitation conditions. This indicates that
the blueshift only depends on the number of polaritons within the
considered state. For instance, when multimode lasing is achieved,
each lasing mode presents its own blueshift, corresponding to its
own occupancy. Thus the blueshift does not come from interaction
with high energy excitons or electron-hole pairs, but mainly from
the self-interaction energy within the considered mode. Notice that
the present results evidence a blueshift logarithmically varying
with the occupancy whereas theoretical calculations predicts a
linear behavior\cite{Ciuti}. Further theoretical investigation is
probably needed to quantitatively describe the observed
self-interaction energy.

To conclude, polariton lasing is demonstrated on the discrete modes
of a GaAs/GaAlAs micropillar cavity. A sharp threshold associated
with a spectral narrowing shows the onset of stimulated scattering
toward the lowest energy polariton state. The polariton state
occupancy is measured to reach $10^{4}$. Changing the excitation
spatial symmetry, multimode polariton lasing is triggered,
demonstrating that the observed feature can not be described in
terms of Bose Einstein condensation. Contrary to photon lasing
obtained at higher excitation power, the polariton self-interaction
within the macroscopically occupied state induces a continuous
blueshift of the emission as the state occupancy builds up. These
results, obtained in the well controlled GaAs semiconductor system,
open the way toward an electrically pumped polariton
laser\cite{APLBajoni} and will stimulate future experiments to
investigate the emission quantum statistics of such a solid-state
matter-wave laser.

We are grateful to D. Le Si Dang for fruitful discussions. This work
was funded by the european project ``Clermont 2"
(MRTN-CT-2003-503677), by "C'nano Ile de France" and "Conseil
G\'en\'eral de l'Essonne".


\begin{thebibliography}{99}
\bibitem{footnote}
Photon-lasing is observed 5 meV below the bare cavity modes deduced
from low density measurements. This effect, pronounced in cavities
containing many QWs, can be understood considering changes of the QW
layer refractive index when screening the exciton.\cite{bajoni}
\end{thebibliography}
\end{document}